\documentclass[twoside,reqno]{qts-proc}

\usepackage[dvips]{graphicx}

\begin{document}
\sloppy \raggedbottom
\setcounter{page}{1}

\newpage
\setcounter{figure}{0}
\setcounter{equation}{0}
\setcounter{footnote}{0}
\setcounter{table}{0}
\setcounter{section}{0}


\title{SIP-potentials and self-similar potentials of
Shabat and Spiridonov: space asymmetric deformation}

\runningheads{Sergei~P.~Maydanyuk and Liliya~M.~Saryan}
{Asymmetry in SIP- and Shabat's and Spiridonov's potentials}

\begin{start}

\author{Sergei~P.~Maydanyuk}{}\thanks{E-mail: maidan@kinr.kiev.ua} %
and \coauthor{Liliya~M.~Saryan}{}

\address{%
Institute for Nuclear Research,\\
National Academy of Sciences of Ukraine\\
prosp. Nauki, 47, Kiev-28, 03680, Ukraine}{}

\begin{Abstract}
An appropriateness of a space asymmetry of shape invariant potentials
with scaling of parameters and potentials of Shabat and Spiridonov
in calculation of their forms, wave functions and discrete energy
spectra has proved and has demonstrated on a simple example.
Parameters, defined space asymmetry, have found.
A new type of a hyerarchy, in which superpotentials with neighbouring
numbers are connected by space rotation relatively a point of origin
of space coordinates, has proposed.
\end{Abstract}
\end{start}



\section{Formalism of shape invariant potentials
\label{sec.1}}

Methods of SUSY QM increase essentially a class of \emph{exactly
solvable potentials}, which can be obtained by \emph{methods of direct
and inverse approaches of quantum mechanics} also. If to assume, that
there is an additional interdependence between potentials-partners
besides of interdependence of SUSY algebra, expressed by Darboux
transformations
(for example, see p.~275--277 in~\cite{Cooper.1995.PRPLC},
p.~4--7 in~\cite{Maydanyuk.2005.APNYA},
\cite{Maydanyuk.2005.Surveys_in_HEP})
or transformations of non-linear supesymmetry
\cite{Andrianov.2004.JPAGB},
then in a number of cases it allows to find a form of such potentials
and their spectral characteristics. If the additional interdependence
points out to a similarity between shapes of these potentials, then
the potentials are named as \emph{shape invariant}
\cite{Gendenshtein.1983.JETPL}.
One can find for them exactly energy spectrum
\cite{Gendenshtein.1983.JETPL} and wave functions
\cite{Dutt.1986.PHLTA}.

We shall name as \emph{shape invariant} (or \emph{SIP-potentials})
the potentials $V_{1}(x)$ and $V_{2}(x)$, if
\begin{equation}
\begin{array}{cc}
  V_{2}(x, a_{1}) = V_{1}(x, a_{2}) + R(a_{1}), &
  R(a_{1}) = const(a_{1}),
\end{array}
\label{eq.1.1}
\end{equation}
where $a_{1}$ is set of parameters, $a_{2}$ is function of $a_{1}$
(see p.~289-290 in~\cite{Cooper.1995.PRPLC}).
Such SIP-potentials have been studied best of all, for which the
dependence of parameters $a_{2}$ on $a_{1}$ has a form of
\emph{translation} (see \cite{Cooper.1987.PHRVA,Dutt.1988.AJPIA})
\begin{equation}
  a_{2} = a_{1} + const,
\label{eq.1.2}
\end{equation}
or \emph{scaling} (see \cite{Barclay.1993.PHRVA,Khare.1993.JPAGB})
\begin{equation}
\begin{array}{ll}
  a_{2} = a_{1} q^{n}, & 0 < q < 1.
\end{array}
\label{eq.1.3}
\end{equation}

In this paper we analyze a possibility of a space asymmetry (at
change $x \to -x$) of the SIP-potentials with scaling at $n=1$ (which
were introduced at the first time in~\cite{Khare.1993.JPAGB}). We
shall assume, that for a superpotential $W(x, a_{1})$ and constant
$R(a_{1})$ there are representations in a form of convergent series
(see p.~299-300 in~\cite{Cooper.1995.PRPLC},
(7)--(8) in~\cite{Khare.1993.JPAGB})
\begin{equation}
\begin{array}{ll}
  W(x, a_{1}) = \sum\limits_{j=0}^{+\infty} g_{j}(x) a_{1}^{j}, &
  R(a_{1})    = \sum\limits_{j=0}^{+\infty} R_{j} a_{1}^{j}.
\end{array}
\label{eq.1.4}
\end{equation}

\section{A general solution for the superpotential
\label{sec.2}}

Using the definition (\ref{eq.1.1}) of SIP-invariancy, Darboux
transformations
\footnote {
In this paper we do not use the nonlinear supersymmetry of
SIP-potentials \cite{Andrianov.2004.JPAGB}},
the representations (\ref{eq.1.4}) and writing down expressions for
variable $a_{1}$ with the same powers, one can make up equations for
obtaining unknown $g_{n}(x)$:
\begin{equation}
\begin{array}{ll}
  2 \displaystyle\frac{d g_{0}(x)}{dx} = R_{0}, &
  \displaystyle\frac{d g_{n}(x)}{dx} + g_{n}(x) f_{n}(x) = h_{n}(x),
\end{array}
\label{eq.2.1}
\end{equation}
\begin{equation}
\begin{array}{ll}
  f_{n}(x) = 2 d_{n} g_{0}(x), &
  h_{n}(x) =
    d_{n} r_{n} - d_{n} \sum\limits_{j=1}^{n-1} g_{j}(x) g_{n-j}(x),
\end{array}
\label{eq.2.2}
\end{equation}
where
\begin{equation}
\begin{array}{lll}
  r_{n} = \displaystyle\frac{R_{n}}{1-q^{n}}, &
  d_{n} = \displaystyle\frac{1-q^{n}}{1+q^{n}}, &
  n = 1, 2, 3 \ldots
\end{array}
\label{eq.2.3}
\end{equation}

A solution of the function $g_{n}(x)$ was found earlier in an
integral form to a constant of integration (for example, see
p.~299--300 in~\cite{Cooper.1995.PRPLC}, in that time compiled a
number of papers at this theme). However, a further construction of
the SIP-potentials was fulfilled usually at zero values of these
constants (also see (12) in~\cite{Khare.1993.JPAGB}).
It turns out, that ``zero'' choise of such constants of integration
reduces the found potentials to their space symmetric solutions
relatively point $x=0$ of origin of space coordinates.
One can conclude, that the SIP-potentials with scaling are
essentially symmetric (see (12) and (13)
in~\cite{Khare.1993.JPAGB}, p.~301 in~\cite{Cooper.1995.PRPLC}).
\emph{Self-similar potentials}, studied by \emph{Shabat}
\cite{Shabat.1992.INPEE} and \emph{Spiridonov}
\cite{Spiridonov.1992.PRLTA}, represent a subset in the set of such
SIP-potentials (see p.~300 in~\cite{Cooper.1995.PRPLC}, p.~4 (5) and
p.~6 in~\cite{Khare.1993.JPAGB}) and should be symmetric also.

However, it turns out, that the non-zero account of the constants of
integration brings a space asymmetry into the SIP-potentials.
Changing these constants, one can deform asymmetrically the
potentials. \emph{Thus, the constants of integration are such unique
parameters, which defines the space asymmetry of the SIP-potentials}.
We do not find any asymmetric forms of the SIP-potentials with scaling
in other papers.

Let's consider a general solution of the function $g_{n}(x)$ (from
(\ref{eq.2.1})):
\begin{equation}
  g_{n}(x) =
    \biggl(\displaystyle\int h_{n}(x)
    e^{\displaystyle\int f_{n}(x) dx} dx + C_{n} \biggr)
    e^{-\displaystyle\int f_{n}(x) dx}.
\label{eq.2.4}
\end{equation}
For the first numbers $n$ we obtain:
\begin{equation}
\begin{array}{lcl}
  g_{0}(x) & = & \displaystyle\frac{R_{0} x}{2} + C_{0}, \\
  g_{1}(x) & = &
    \biggl(\displaystyle\int d_{1} r_{1}
    e^{\displaystyle\frac{d_{1} R_{0} x^{2}}{2} + 2 d_{1}C_{0} x}
    dx + C_{1} \biggr)
    e^{-\displaystyle\frac{d_{1} R_{0} x^{2}}{2} - 2 d_{1}C_{0} x}, \\
  g_{n}(x) & = &
    \biggl(\displaystyle\int d_{n} \biggl(r_{n} -
      \sum\limits_{j=1}^{n-1} g_{j}(x) g_{n-j}(x) \biggr)
    e^{\displaystyle\frac{d_{n} R_{0} x^{2}}{2} + 2 d_{n}C_{0} x}
    dx \; + \\
  & & + \;
    C_{n} \biggr)
    e^{-\displaystyle\frac{d_{n} R_{0} x^{2}}{2} - 2 d_{n}C_{0} x},
\end{array}
\label{eq.2.5}
\end{equation}
where $C_{n}$ are the constants of integration.

We see, that the found functions $g_{n}(x)$ at non-zero $C_{n}$ are
space asymmetric relatively point $x=0$ and give the asymmetric
solutions for the superpotential and potentials-partners, connected
with it. The solution (\ref{eq.2.5}) is the \emph{general} one and
it can be considered as the asymmetric generalization of the known
solutions (160) in~\cite{Cooper.1995.PRPLC}, (12)
in~\cite{Khare.1993.JPAGB} (which in a context of these papers are
antisymmetric).

One can see, that the definition of the SIP-potentials determines the
function $g_{n}(x)$ to $n$ arbitrary constants $C_{i}$
($i=1 \ldots n$). For arbitrary values of the constants $C_{n}$ the
function $g_{n}(x)$ in the form (\ref{eq.2.5}) is the solution of the
system (\ref{eq.2.1}). \emph{Therefore, the possibility of difference
of the constants $C_{n}$ on their zero values is rightful, the
possibility of the asymmetry of the SIP-potentials is rightful}.

We see, that an approach for calculation of the discrete component of
the energy spectrum of the SIP-potentials with scaling of the
parameters (see (173), p.~302 in~\cite{Cooper.1995.PRPLC}; solutions
(171) in~\cite{Cooper.1995.PRPLC}) has an arbitrariness in selection
of the values for the constants $C_{n}$. \emph{Therefore, the
possibility of difference of the constants $C_{n}$ on their zero
values in calculation of energy spectra is rightful}.

\section{Are the potentials of Shabat and Spiridonov symmetric?
\label{sec.3}}

The potentials, studied by Shabat and Spiridonov, belong to a set of
the SIP-potentials with scaling of the parameters and, because of
this, they can be space asymmetric also. However, usually an
independent formalism is used for their description
(see \cite{Spiridonov.1992.PRLTA,Shabat.1992.INPEE,Barclay.1993.PHRVA}).
Further, we shall show, that on the basis of such a formalism one can
obtain the asymmetry of these potentials also.

Let's define superpotentials of Shabat and Spiridonov by such way
(see (8) in~\cite{Spiridonov.1992.PRLTA};
(2.8) in~\cite{Barclay.1993.PHRVA}):
\begin{equation}
  W_{n}(x) = q^{n} W(q^{n} x).
\label{eq.3.1}
\end{equation}
We shall use a restriction $0<q<1$, though, according to
\cite{Spiridonov.1992.PRLTA}, it is not obligatory (as against
\cite{Shabat.1992.INPEE}).
Then using interdependence between the superpotentials with
neighboring numbers of an arbitrary hierarchy (for example, see (2.4)
in~\cite{Barclay.1993.PHRVA}):
\begin{equation}
\begin{array}{ll}
  W_{n}^{2}(x) + \displaystyle\frac{d W_{n}(x)}{dx} =
    W_{n+1}^{2}(x) - \displaystyle\frac{d W_{n+1}(x)}{dx} + k_{n+1}, &
  n = 0, 1, 2...
\end{array}
\label{eq.3.2}
\end{equation}
(constants $k_{n}$ are defined by distances between neighboring
levels $E_{n+1}^{(0)}$ and $E_{n}^{(0)}$ of the first potential
$V_{0}(x)$) and using expansion of the function $W(x)$ in series near
to zero point $x=0$:
\begin{equation}
\begin{array}{cc}
  W(x) = \sum\limits_{m=0}^{+\infty} b_{m} x^{m}, &
  b_{m} = const,
\end{array}
\label{eq.3.3}
\end{equation}
one can find relations between $b_{m}$ and $k_{n}$ ($n = 0,1,2 \dots$):
\begin{equation}
\begin{array}{l}
  q^{2n} (b_{1} (1+q^{2}) + b_{0}^{2} (1-q^{2})) = k_{n+1}, \\
  b_{m+1} = -\displaystyle\frac{1-q^{m+2}}{(m+1)(1+q^{m+2})}
            \sum\limits_{i=0}^{m} b_{i} b_{m-i}.
\end{array}
\label{eq.3.4}
\end{equation}

Let's calculate the first coefficients $b_{n}$:
\begin{equation}
\begin{array}{lcl}
  b_{1} & = & \displaystyle\frac
              {k_{1} - b_{0}^{2}(1-q^{2})} {1+q^{2}}, \\

  b_{2} & = & -\displaystyle\frac{1-q^{3}}{1+q^{3}} b_{0} b_{1} =
              -\displaystyle\frac{1-q^{3}}{(1+q^{3})(1+q^{2})} b_{0}
              \Bigl(k_{1} - b_{0}^{2}(1-q^{2}) \Bigr), \\

  b_{3} & = &
              -\displaystyle\frac{1-q^{2}}{3(1+q^{4})}
              \Bigl(k_{1} - b_{0}^{2}(1-q^{2}) \Bigr)
              \biggl( -2b_{0}^{2} \displaystyle\frac{1-q^{3}}{1+q^{3}}
                     +\displaystyle\frac{k_{1} - b_{0}^{2}(1-q^{2})}
                     {1+q^{2}} \biggl).
\end{array}
\label{eq.3.5}
\end{equation}
One can see, that at $b_{0}=0$ all coefficients $b_{n}$ with even
numbers $n$ are equal to zero, the function $W(x)$ are antysymmetric
and coincides with the known solution (2.17)
in~\cite{Barclay.1993.PHRVA}. Changing the values of the coefficients
$b_{0}$ and $k_{1}$, one can deform the shape of the function $W(x)$,
obtaining new asymmetric solutions for the potentials.

\begin{figure}[ht]
\centerline{
\includegraphics[width=5cm]{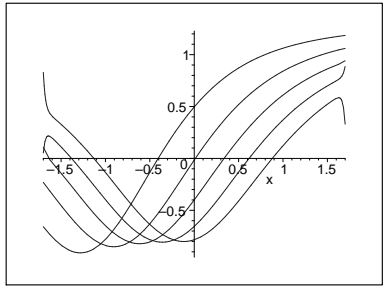}}
\caption{Displacement of the potential at
$b_{0}$ = 0.1, 0.3, 0.5, 0.7, 0.9
\label{fig.422}}
\end{figure}

\underline{Example}.
In Fig.~\ref{fig.422} a displacement of the potential shape with
changing of $b_{0}$ has shown ($k_{1} = 1.0$, $q = 0.5$; for
calculations of the function $W(x)$ a partial sum $W^{(n)}(x)$
at $n=75$ is used).
For estimation of applicability of the method, let's find a region of
convergence, using a definition of the partial sum:
\begin{equation}
  W^{(n)}(x) = \sum\limits_{j=0}^{n} b_{j} x^{j}.
\label{eq.3.6}
\end{equation}
Values of these sums in selected points $x$ and at selected $n$
are presented in the table (at $b_{0} = 0.$):
\begin{center}
\begin{tabular}{|c|c|c|c|} \hline
       &     $x = 1.0$     &     $x = 1.5$     &    $x = 2.0$     \\ \hline
n= 1   & 0.800009999940000 & 1.20000999991000  & 1.60000 \\ \hline
n= 5   & 0.670153606255581 & 1.00805575100451  & 1.96237 \\ \hline
n=10   & 0.657567324973884 & 0.914274086501970 & 2.53156 \\ \hline
n=20   & 0.656219813282069 & 0.809852235822209 &-3.76793 \\ \hline
n=50   & 0.656223635811415 & 0.824633977282796 & 116.677 \\ \hline
n=100  & 0.656223635811327 & 0.824569097466698 &-24219.3 \\ \hline
n=200  & 0.656223635811327 & 0.824569105154386 &-1.06035 $10^{+9}$ \\ \hline
n=500  & 0.656223635811327 & 0.824569105154386 &-8.89752 $10^{+22}$ \\ \hline
\end{tabular}
\end{center}
One can see, that in the region $|x|<1.5$ the partial sums $W_{n}(x)$
converge to the function $W(x)$ easily enough. However, at values
of $|x|$ more then 1.5 (approximately) it is not possible to achieve
such a convergence. With change of the parameters $b_{0}$, $q$ and
$k_{1}$ (up to 1) the convergence of the function $W(x)$ is kept.
Therefore, one can accept the region $|x|<1.5$ as the \emph{region
of the convergence} of the representations (\ref{eq.3.3}) for the
function $W(x)$, \emph{within which limits one can accept the shape
of the function $W(x)$ and its asymmetric deformation as rightful}.

\section{A hierarhy with rotations
\label{sec.4}}

Interesting type of deformation of the superpotential $W_{n}(x)$ is
its rotation in a plane $(x, W_{n}(x))$ around point of origin to a
given angle.

\subsection{Operator of rotation 
\label{sec.4.1}}

Let's consider a point $A$ of a function $f(x)$ with coordinates 
$x_{A}$ and $f_{A}=f(x_{A})$.
We introduce operator of rotation:
\emph{operator $T_{\Delta\varphi}$ rotates any point of the function 
$f(x)$ in a plane $(x, f(x))$ around point of origin to an angle
$\Delta \varphi$:}
\begin{equation}
  T_{\Delta\varphi} f_{old} (x_{old}) = f_{new} (x_{new}).
\label{eq.4.1.1}
\end{equation}

Introducing a complex function $z(x)$ by such a way:
\begin{equation}
\begin{array}{ll}
  Re (z) = x; &
  Im (z) = f(x),
\end{array}
\label{eq.4.1.2}
\end{equation}
one can describe the rotation of the function $f(x)$:
\begin{equation}
\begin{array}{cc}
  z_{old} = \rho e^{i\varphi_{old}}, &
  z_{new} =
    \rho e^{i\varphi_{new}} =
    T_{\Delta\varphi} \rho e^{i\varphi_{old}} =
    \rho e^{i(\varphi_{old} + \Delta\varphi)},
\end{array}
\label{eq.4.1.3}
\end{equation}
and also one can find a view of the operator of the rotation:
\begin{equation}
\begin{array}{cc}
  T_{\Delta\varphi} = e^{i\Delta\varphi}, &
  \Delta\varphi = \varphi_{new} - \varphi_{old}.
\end{array}
\label{eq.4.1.4}
\end{equation}
Then we obtain transformations of coordinates in the rotation:
\begin{equation}
\begin{array}{l}
  x_{new} = Re (z_{new}) =
    Re (\rho e^{i(\varphi_{old} + \Delta\varphi)}) =
    x_{old} \cos{\Delta\varphi} - f_{old} \sin{\Delta\varphi}, \\
  f_{new} = Im (z_{new}) =
    Im (\rho e^{i(\varphi_{old} + \Delta\varphi)}) =
    f_{old} \cos{\Delta\varphi} + x_{old} \sin{\Delta\varphi}.
\end{array}
\label{eq.4.1.5}
\end{equation}

\subsection{Construction of a hierarchy of superpotentials with use
of the rotations
\label{sec.4.2}}

Let's use the function $f_{n}(x)$ with number $n$ for description of
the superpotential $W_{n}(x)$ with number $n$. Then a sequence of the
functions $f_{0}(x)$, $f_{1}(x)$ ... represents a hierarchy of the
superpotentials $W_{n}(x)$, which can be transformed one to another
by use of such rotations.

We assume, that one can write the superpotential with the number $n$
in some vicinity of point $x=x_{n}$ in a form of the power series:
\begin{equation}
  W_{n} (x) = \sum\limits_{m=0}^{+\infty} a_{m}^{(n)} x^{m}.
\label{eq.4.2.1}
\end{equation}
For this superpotential at point $x=x_{n}$ we find:
\begin{equation}
\begin{array}{l}
  \displaystyle\frac{d W_{n}(x)}{dx} \biggr|_{x=x_{n}} =
    \sum\limits_{m=0}^{+\infty} (m+1) a_{m+1}^{(n)} x^{m}_{n}, \\

  W_{n}^{2} (x_{n}) =
    \sum\limits_{m=0}^{+\infty} \sum\limits_{k=0}^{+\infty}
    a_{m}^{(n)} a_{k}^{(n)} x^{m+k}_{n} \equiv
    \sum\limits_{m=0}^{+\infty} a_{(2),m}^{(n)} x^{m}_{n}.
\end{array}
\label{eq.4.2.2}
\end{equation}

Taking into account the coordinate transformations (\ref{eq.4.1.5})
by rotation and representation (\ref{eq.4.2.1}), we obtain:
\begin{equation}
\begin{array}{l}
  x_{n+1} =
    \cos{\Delta\varphi} x_{n} -
    \sin{\Delta\varphi}
    \sum\limits_{m=0}^{+\infty} a_{m}^{(n)} x^{m}_{n}, \\

  x_{n-1} =
    x_{n} - \Delta x_{n} =
    (2 - \cos{\Delta\varphi}) x_{n} +
    \sin{\Delta\varphi}
    \sum\limits_{m=0}^{+\infty} a_{m}^{(n)} x^{m}_{n} \equiv
    \sum\limits_{m=0}^{+\infty} \bar{a}_{m}^{(n)} x^{m}_{n},
\end{array}
\label{eq.4.2.3}
\end{equation}
where $\Delta x_{n} = x_{n+1} - x_{n}$.

From here, one can calculate the superpotential $W_{n+1}(x)$ at point
$x_{n}$:
\begin{equation}
\begin{array}{lcl}
  W_{n+1} (x_{n}) & = &
    \cos{\Delta\varphi}
    \sum\limits_{m=0}^{+\infty} a_{m}^{(n)} x^{m}_{n-1} +
    \sin{(\Delta\varphi)} x_{n-1} \equiv
    \sum\limits_{m=0}^{+\infty} \bar{a}_{m}^{(n+1)} x^{m}_{n}.
\end{array}
\label{eq.4.2.4}
\end{equation}
And we find:
\begin{equation}
\begin{array}{l}
  \displaystyle\frac{d W_{n+1}(x)}{dx} \biggr|_{x=x_{n}} =
    \sum\limits_{m=0}^{+\infty} (m+1) \bar{a}_{m+1}^{(n+1)} x^{m}_{n}, \\

  W_{n+1}^{2} (x_{n}) =
    \sum\limits_{m=0}^{+\infty} \sum\limits_{k=0}^{+\infty}
    \bar{a}_{m}^{(n+1)} \bar{a}_{k}^{(n+1)} x^{m+k}_{n} \equiv
    \sum\limits_{m=0}^{+\infty} \bar{a}_{(2),m}^{(n+1)} x^{m}_{n}.
\end{array}
\label{eq.4.2.5}
\end{equation}

Using (\ref{eq.4.2.2}) and (\ref{eq.4.2.5}), and also expressions
(\ref{eq.3.2}) of the interdependence between the superpotentials
with the neighboring numbers, we obtain the following system of
equations for calculation of unknown coefficients $a_{m}^{(n)}$:
\begin{equation}
\begin{array}{ll}
  a_{(2),0}^{(n)} + a_{1}^{(n)} -
    \bar{a}_{(2),0}^{(n+1)} + \bar{a}_{1}^{(n+1)} = k_{n+1}, & \\
  a_{(2),m}^{(n)} + (m+1) a_{m+1}^{(n)} -
    \bar{a}_{(2),m}^{(n+1)} + (m+1) \bar{a}_{m+1}^{(n+1)} = 0, 
    \mbox{ при } m \ge 1.
\end{array}
\label{eq.4.2.6}
\end{equation}
Resolving the system (\ref{eq.4.2.6}) relatively the coefficients
$a_{m}^{(n)}$, from (\ref{eq.4.2.1}) we obtain the shape of the
superpotential $W_{n}(x)$ with the number $n$ in the vicinity of
point $x=x_{n}$. If to construct the new superpotential $W_{n+1}(x)$
with the next number $n+1$ by use of the rotation of this
superpotential with the number $n$, then they should be connected
by the SUSY transformations (\ref{eq.3.2}).

By such a way one can construct whole hierarchy of the superpotentials
with rotations. However, here it needs else to use requirement, that
the arbitrary consequence of the rotations of a superpotential with a
given number $n$ gives a new superpotential, which must be determined
uniquely on the whole region of $x$. For example, one can use a
convergent geometric series for such a consequence of rotations.

\section{Conclusions
\label{sec.conclusions}}

In finishing we note main conclusions and new results.
\begin{itemize}

\item
An appropriateness of a space asymmetry of shape invariant
potentials with scaling of parameters
in calculation of their forms, wave functions and discrete energy
spectra has proved.

\item
The coefficients $C_{n}$ are such unique parameters, which determine
the space asymmetry of the SIP-potentials (without displacement of
levels in spectra).

\item
It has shown, that the definition of the potentials of Shabat and
Spiridonov in the form (\ref{eq.3.2}) allows their space asymmetric
deformation.

\item
The general solution for the function $W(x)$ and the superpotential
$W_{n}(x)$ of Shabat and Spiridonov with number $n$ are defined
by three parameters: $q$, $k_{1}$ and $b_{0}$.
The parameter $q$ allows to deform a shape of the potentials, not
influencing to their asymmetry.
Energy spectrum is determined uniquely by two parameters $q$ and
$k_{1}$ and is not depended on $b_{0}$.
The parameter $b_{0}$ is only one parameter, determining the
asymmetry of the potentials.

\item
A new approach for construction of the hierarchy, in which the
superpotentials with the neighboring numbers are connected by their
rotation relatively point of origin, has proposed.

\end{itemize}

\section*{Acknowledgments}

One of authors expresses his deep gratitude to the organizers of the
4th International Symposium \emph{``Quantum Theory and Symmetries''}
for warm hospitality,
to members of EWTF group of European Physical Society and
to members of Ukrainian Physical Society for
grant for supporting our participation in this nice meeting.

\bibliography{QTS4}

\end{document}